\newcommand{\paperversion}{final}
\providecommand{\paperversion}{submission}
\providecommand{\anonymoussubmission}{1}
\RequirePackage{paperversions}

\documentclass[sigconf, screen, pbalance, dvipsnames, language]{acmart}

\usepackage[utf8]{inputenc}

\usepackage{booktabs}   \usepackage{subcaption} 

\usepackage{amsmath}
\usepackage{mathtools}
\mathtoolsset{mathic}
\usepackage{mathpartir}
\usepackage{pl-syntax}

\usepackage{cleveref}
  \crefformat{section}{\S#2#1#3}
  \crefname{lstlisting}{listing}{listings}
  \Crefname{lstlisting}{Listing}{Listings}

\usepackage[pages=3]{checkpagelimit}
\ifshowcomments
  \usepackage{aplcomments}
\else
  \usepackage[disabled]{aplcomments}
\fi

\usepackage{defs}
\usepackage{abstract-syntax}
\usepackage{concrete-syntax}
\usepackage{meta-variables}
\usepackage{information-flow}

\setcitestyle{nosort}

\setcopyright{none}
\acmPrice{}
\acmDOI{}
\acmYear{2023}
\copyrightyear{2023}
\acmSubmissionID{}
\acmISBN{}
\acmConference
  [FCS 2023]
  {Workshop on Foundations of Computer Security}
  {July 9, 2023}
  {Dubrovnik, Croatia}
\acmBooktitle{}

\newcommand{\cornell}{
  \affiliation{\institution{Cornell University}
\city{Ithaca}
    \state{NY}
    \postcode{14850}
    \country{USA}
  }
}

\author{Vivian Ding}
\orcid{0009-0008-5110-0593}
\cornell
\email{vyd2@cornell.edu}

\author{Co\c{s}ku Acay}
\orcid{0000-0002-0487-1167}
\cornell
\email{cacay@cs.cornell.edu}

\author{Andrew C. Myers}
\orcid{0000-0001-5819-7588}
\cornell
\email{andru@cs.cornell.edu}
 \title{An Array Intermediate Language for Mixed Cryptography}

\begin{document}

\maketitle

\section{Introduction}
\label{sec:intro}

Advanced cryptographic mechanisms such as secure multiparty
computation (MPC)~\cite{yao82}, zero-knowledge proofs (ZKP)~\cite{GoldwasserMR89},
and fully homomorphic encryption (FHE)~\cite{fhe} all expect programs
represented as fixed-sized circuits.
Since it is painful to program circuits directly,
cryptographic compilers~\cite{fairplay, efficient-mpc,ScaleMamba, wysteria,
oblivm,pinocchio,geppetto,buffet,viaduct-pldi21, LevySIKMZ23, CHET,EVA,
porcupine,heco,coyote,gazelle}
alleviate the burden on programmers by translating high-level code to low-level
circuits.
However, many existing compilers fail to take full advantage of the underlying
cryptographic libraries to speed up execution, as neither the
source representation nor the circuit representation are suitable for
optimizations such as vectorization.

Prior work on cryptographic intermediate representations (IRs), including that
focusing on vectorization for cryptographic mechanisms~\cite{LevySIKMZ23, OzdemirBW22}, targets programs whose entire behavior can be implemented as a single
cryptographic circuit.
This approach therefore cannot support interactive programs in which user input arrives
during execution; nor does it support the compilation of programs mixing
multiple cryptographic back ends and local computation, which is useful for
performance.

We introduce \ir, a new array-based intermediate representation designed to
support generating efficient code for interactive programs employing multiple
cryptographic mechanisms.
\ir is intended as an IR for the Viaduct
compiler~\cite{viaduct-pldi21}, which can synthesize secure, distributed
programs with an extensible suite of cryptography.
Therefore, \ir supports an extensible variety of cryptographic
mechanisms, including MPC and ZKP.
It is the job of the Viaduct compiler to select cryptographic protocols
that make the compiled program secure, guided by information-flow
annotations. In this paper, we assume this choice has been made
correctly, and focus on the IR. A proof that Viaduct generates secure
target code in the sense of UC simulation~\cite{UC}, assuming that cryptographic
mechanisms are described faithfully by the information-flow
annotations, is provided elsewhere~\cite{AGRM23}.

\begin{lstlisting}[
    float,
    caption={An interactive biometric matching service.},
    label={fig:biometric}
  ]
host Client, Server

circuit fun <n, d> biometric@MPC(
  database: int[n, d],
  sample: int[d]) -> min_dist: int[] {
    val dists[i < n] = reduce(::+, 0) { j < d ->
        (database[i, j] - sample[j]) ** 2
    }
    val res[] = reduce(::min, int.MAX) { i < n ->
        dists[i]
    }
    return res
}

fun main() {
    val N@Replication(Client, Server) = 1000
    val D@Replication(Client, Server) = 20
    val database@Local(Server) =
        Server.input<int[N, D]>()
    while (true) {
        val sample@Local(Client) =
            Client.input<int[D]>()
        val result@Local(Client) =
            biometric<N, D>(database, sample)
        Client.output(result)
    }
}
\end{lstlisting}

\section{Overview of \ir}
Large-scale applications may require multiple cryptographic mechanisms at
different points of a program, as well as reactive functionality or running
computations indefinitely. \ir represents code employing multiple cryptographic
mechanisms as structured control flow over calls to circuit functions.
To demonstrate how our approach enables these programs, we
use a program for interactive biometric matching as a running example.
In \cref{fig:biometric}, the circuit function
\lstinline{biometric} describes a computation to be performed on \lstinline{MPC},
which represents a secure multiparty computation protocol between
\lstinline{Server} and \lstinline{Client}.
It matches a \lstinline{d}-tuple sample against a database of \lstinline{n}
points and reports the squared Euclidean distance between the sample and the
closest database entry.
The result is a zero-dimensional array (a scalar), as indicated by the syntax
\lstinline{int[]}.

The function \lstinline{main} performs an infinite loop of inputs, outputs,
and calls to the circuit function \lstinline{biometric}.
Results of function calls are bound to variables,
which are associated with \emph{storage formats} describing how the variables
are stored across hosts.
In the example, values specifying array sizes are stored using a replication
protocol, as both parties must have access to array sizes.
The input samples and results of calling \lstinline{biometric} are stored
locally by \lstinline{Client}.

\section{Syntax}
\begin{figure}
  \begin{syntax}
  \groupleft{
    \categoryFromSet[Hosts]{\h}{\h!}
    \hfill
    \categoryFromSet[Computation Protocol]{\cp}{\cp!}
  }
  \groupleft{
    \categoryFromSet[Variables]{\x}{\x!}
    \hfill
    \categoryFromSet[Storage Protocol]{\stp}{\stp!}
  }

  \separate

  \categoryFromSet[Values]{\val}{\val!}
  \categoryFromSet[Types]{\type}{\type!}
  \categoryFromSet[Function Names]{\f}{\f!}

  \separate

  \category[Atomic Expr.]{\aexp}
  \alternative{\val}
  \alternative{\x}

  \category[Index Bounds]{\ib}
  \alternative{\x < \aexp}

  \category[Scalar Expr.]{\scexp}
  \alternative{\elookup{\x}{\many{\aexp}}}
  \alternative{\eapplyop{\op}{\scexp_1, \scexp_2}}
  \\
  \alternative{\ereduce{\op}{\scexp}{\ib}{\scexp}}

  \category[Commands]{\m}
  \alternative{\aexp}
  \alternative{\ecall{\f}{\many{\aexp}}{\many{\aexp}}}
  \\
  \alternative{\einput{\h}{\type}}
  \alternative{\eoutput{\h}{\aexp}}

  \separate

  \category[Circuit Statements]{\cs}
  \alternative{\slet{\elookup{\x}{\many{\ib}}}{\scexp}}

  \category[Statements]{\s}
  \alternative{\sletp{\x}{\stp}{\m}}
  \\
  \alternative{\sif{\aexp}{\s_1}{\s_2}}

  \category[Parameters]{\param}
  \alternative{\x : \type}

  \category[Top-level Decl.]{\decl}
  \groupleft{
    \alternative{
      \dcircfun{\many{\x}}{\f}{\cp}{\many{\param}}{\many{\param}}{\many{\cs}}{\many{\x}}
    }
  }
  \groupleft{
    \alternative{
      \dfun{\many{\x}}{\f}{\many{\param}}{\many{\param}}{\many{\s}}{\many{\x}}
    }
  }
\end{syntax}
   \caption{Syntax of \ir.}
  \label{fig:syntax}
\end{figure}

\Cref{fig:syntax} gives the syntax of \ir.
Circuit functions are straight-line blocks of computations performed on a
single cryptographic protocol, parameterized by array sizes.
Computations in circuit functions are expressed in the form of
assignments to multidimensional arrays.
The expressions used in these assignments include arithmetic and logical
operations, indexing into arrays, references to named values (including size
parameters), and bulk operations such as \lstinline{reduce}.
While each circuit function must be associated with one protocol,
a program may employ multiple protocols across circuit functions.
Non-circuit functions contain control-flow statements (conditionals) and
let-bindings, which bind constants, inputs, outputs, and the results
of circuit or function calls to named values. We elide looping constructs in
the formal language since they can be recovered through recursion. No
computation can occur in non-circuit functions.

\section{Array-Based Computation}
It is a standard restriction in cryptographic mechanisms such as MPC that
bounds must be known at the time of circuit generation.
Cryptographic circuits require all loops to be unrolled, conditional statements
to be translated into ``muxes'', and functions to be inlined~\cite{fairplay}.
However, this process blows up the program size and erases all structure in the resulting
circuit representation, making it difficult and expensive to perform optimizations.

Array programs are conditional and loop-free, and are therefore a useful
representation for translation to low-level circuits, unlike high-level
languages. Many loops in practice can be expressed with array operations, so
programs do not grow unreasonably in size.
Array programs also preserve structure and natively capture bulk operations,
so they are easy to optimize and parallelize, unlike circuits.
As vectorization is a crucial optimization for efficiency
~\cite{LevySIKMZ23, motion, porcupine, heco, coyote},
this array-centric representation may present significant
speedup in generated code.

As such, computations in \ir are expressed solely using arrays.
Circuit functions in \ir are parameterized over sizes to support protocols
which require bounds to be known at generation time.

\section{Control Flow}
Additionally, \ir separates control flow from cryptographic execution by
restricting all computation to pure circuit blocks. Inputs, outputs, calls to
circuits, and other control-flow statements occur only in non-circuit functions.
This representation prevents arbitrary control flow inside of circuits, without
restricting the overall program to straight-line computations.
It also enables reactivity: programs can exit from circuit functions to
perform input and outputs before resuming computation.

Thus, in the biometric example, while the size of the circuit is bounded, the
interactive loop between the client and server can continue indefinitely.

\section{Protocol Mixing} 

Intermediate storage may be needed if values computed by a circuit function are
used in other circuits.
For some cryptographic implementations such as ABY~\cite{aby}, circuits are
destroyed upon evaluation. In order to support intermediate results, values must
be explicitly exported and imported as secret shares.

In \ir, cryptographic protocols may be used for computation or intermediate
storage.
For instance, commitment schemes and secret sharing are storage protocols, as
they specify a format for storing data. ZKP and MPC are computation protocols,
and replication is both a storage and a computation protocol.
Each protocol is associated with data formats for storing or computing values.
To execute circuit functions, data must be imported and exported between these
formats.

In the biometric example, the call to \lstinline{biometric} is associated with
two imports (\lstinline{database} and \lstinline{sample}, imported from local
cleartext formats to the format for computation with MPC) and one export
(\lstinline{result}, exported from the MPC computation format to the client’s
local storage).

Protocol back ends define how both to \emph{import} values to the format
used for cryptographic execution and to \emph{export} data to intermediate
storage.
For instance, to export local values to the replicated format (in which
data is replicated on all hosts involved with the protocol), hosts involved
in replication must perform equivocation checks to ensure that the value they
receive is the same as other hosts in the replication protocol.
Libraries for MPC such as ABY~\cite{aby} allow for execution of circuits in
multiple schemes (arithmetic sharing, boolean sharing, and Yao's garbled
circuits), and ABY supports transferring data between them.

Thus, defining communication between protocols in terms of transfers between
storage and execution formats provides a useful abstraction
for interactive computation that mixes cryptography.

\section{Generation of \ir}

Circuit functions provide cryptographic back ends with large, contiguous blocks
of straight-line computation as opposed to individual instructions.
Working with blocks exposes parallelism opportunities and enables more
advanced optimizations such as algebraic rewrites to simplify
computations~\cite{egg}.
Additionally, large blocks amortize the cost of
interfacing with cryptographic frameworks:
building, evaluating, and destroying circuits,
and importing and exporting data.
Therefore, generating an IR which packs as much computation as possible
into each circuit function is advantageous for efficiency.

We propose to automatically \emph{split} the input program into
(large) circuit functions.
The input to splitting is an array program which freely mixes computation
and control flow, and where each statement is annotated with the
protocol executing that statement.\footnote{
  Protocol annotations can be provided by the programmer or inferred
  automatically through information-flow analysis~\cite{viaduct-pldi21,zcmz03,swift07}.
}
The splitting procedure attempts to group together statements on the
same protocol by reordering them. The goal of reordering is to maximize block
size and to minimize data dependence between blocks (which becomes
communication in the form of imports and exports in the generated IR).

The compiler cannot freely reorder statements: in addition to being restricted
by data dependencies, some reorderings violate security.
The compiler cannot move an \lstinline{output} statement before an
\lstinline{input} statement, and it cannot move a statement that
reveals information before one that commits to data.
For example, if the source program states \lstinline{Client} guesses
a number and then \lstinline{Server} reveals the secret number it picked,
the compiler should not switch the order of these statements.
These constraints precisely identify when it is safe to reorder~\cite{AGRM23}.

\section{Conclusion}

We present a new intermediate representation for compilers that generate
code for advanced cryptographic libraries.
Our IR supports interactive programs by making explicit the boundary between
control flow and cryptographic computation.
It allows mixing local computation with multiple different cryptographic
mechanisms by distinguishing storage formats from computation protocols.
Finally, it facilitates vectorization and other optimizations by partitioning
source programs into large contiguous blocks of array programs.
Fully integrating the new IR into \sysname is a work in progress.
Still to be completed is implementing splitting, and taking full advantage of
vectorization.
We have integrated replication and MPC (through ABY~\cite{aby}),
and hope to integrate ZKP and fully homomorphic encryption (FHE).

\finalpage

\bibliography{packages/bibtex/pm-master}

\newcommand{\etalchar}[1]{$^{#1}$}
\begin{thebibliography}{WNW{\etalchar{+}}21}

\bibitem[ACK{\etalchar{+}}19]{ScaleMamba}
Abdelrahaman Aly, Daniele Cozzo, Marcel Keller, Emmanuela Orsini, Dragos
  Rotaru, Peter Scholl, Nigel~P. Smart, and Tim Wood.
\newblock {\em {SCALE–MAMBA} {v}1.6 : Documentation}, 2019.

\bibitem[AGRM23]{AGRM23}
Coşku Acay, Joshua Gancher, Rolph Recto, and Andrew Myers.
\newblock Secure synthesis of distributed cryptographic applications.
\newblock In submission, 2023.

\bibitem[ARG{\etalchar{+}}21]{viaduct-pldi21}
Coşku Acay, Rolph Recto, Joshua Gancher, Andrew Myers, and Elaine Shi.
\newblock Viaduct: An extensible, optimizing compiler for secure distributed
  programs.
\newblock In {\em 42\textsuperscript{nd} {ACM SIGPLAN} Conf.~on Programming
  Language Design and Implementation (PLDI)}, pages 740--755. {ACM}, June 2021.

\bibitem[BDST22]{motion}
Lennart Braun, Daniel Demmler, Thomas Schneider, and Oleksandr Tkachenko.
\newblock Motion – a framework for mixed-protocol multi-party computation.
\newblock {\em ACM Trans. Priv. Secur.}, 25(2), 2022.

\bibitem[Can01]{UC}
Ran Canetti.
\newblock Universally composable security: {A} new paradigm for cryptographic
  protocols.
\newblock In {\em 42\textsuperscript{nd} Symposium on Foundations of Computer
  Science {(FOCS)}}, pages 136--145. {IEEE} Computer Society, 2001.

\bibitem[CDA{\etalchar{+}}21]{porcupine}
Meghan Cowan, Deeksha Dangwal, Armin Alaghi, Caroline Trippel, Vincent~T. Lee,
  and Brandon Reagen.
\newblock Porcupine: A synthesizing compiler for vectorized homomorphic
  encryption.
\newblock In {\em 42\textsuperscript{nd} {ACM SIGPLAN} Conf.~on Programming
  Language Design and Implementation (PLDI)}, pages 375--389, 2021.

\bibitem[CFH{\etalchar{+}}15]{geppetto}
Craig Costello, C{\'e}dric Fournet, Jon Howell, Markulf Kohlweiss, Benjamin
  Kreuter, Michael Naehrig, Bryan Parno, and Samee Zahur.
\newblock Geppetto: Versatile verifiable computation.
\newblock In {\em IEEE Symp.~on Security and Privacy}, pages 253--270. IEEE,
  2015.

\bibitem[CLM{\etalchar{+}}07]{swift07}
Stephen Chong, Jed Liu, Andrew~C. Myers, Xin Qi, K.~Vikram, Lantian Zheng, and
  Xin Zheng.
\newblock Secure web applications via automatic partitioning.
\newblock In {\em 21\textsuperscript{st} {ACM} Symp.~on Operating System
  Principles (SOSP)}, pages 31--44, October 2007.

\bibitem[DKS{\etalchar{+}}20]{EVA}
Roshan Dathathri, Blagovesta Kostova, Olli Saarikivi, Wei Dai, Kim Laine, and
  Madan Musuvathi.
\newblock {EVA:} an encrypted vector arithmetic language and compiler for
  efficient homomorphic computation.
\newblock In Alastair~F. Donaldson and Emina Torlak, editors, {\em
  41\textsuperscript{st} {ACM SIGPLAN} Conf.~on Programming Language Design and
  Implementation (PLDI)}, pages 546--561. {ACM}, 2020.

\bibitem[DSC{\etalchar{+}}19]{CHET}
Roshan Dathathri, Olli Saarikivi, Hao Chen, Kim Laine, Kristin~E. Lauter, Saeed
  Maleki, Madanlal Musuvathi, and Todd Mytkowicz.
\newblock {CHET:} an optimizing compiler for fully-homomorphic neural-network
  inferencing.
\newblock In Kathryn~S. McKinley and Kathleen Fisher, editors, {\em
  40\textsuperscript{th} {ACM SIGPLAN} Conf.~on Programming Language Design and
  Implementation (PLDI)}, pages 142--156. {ACM}, 2019.

\bibitem[DSZ15]{aby}
Daniel Demmler, Thomas Schneider, and Michael Zohner.
\newblock {ABY} - {A} framework for efficient mixed-protocol secure two-party
  computation.
\newblock In {\em Network and Distributed System Security Symp.} The Internet
  Society, 2015.

\bibitem[Gen09]{fhe}
Craig Gentry.
\newblock Fully homomorphic encryption using ideal lattices.
\newblock In {\em 41\textsuperscript{st} {ACM} Symp. on Theory of Computing},
  pages 169--178, 2009.

\bibitem[GMR89]{GoldwasserMR89}
Shafi Goldwasser, Silvio Micali, and Charles Rackoff.
\newblock The knowledge complexity of interactive proof systems.
\newblock {\em {SIAM} J. Comput.}, 18(1):186--208, 1989.

\bibitem[IMZ19]{efficient-mpc}
Muhammad Ishaq, Ana Milanova, and Vassilis Zikas.
\newblock Efficient {MPC} via program analysis: {A} framework for efficient
  optimal mixing.
\newblock In {\em 26\textsuperscript{th} ACM Conf.\@~on Computer and
  Communications Security (CCS)}, pages 1539--1556. {ACM}, 2019.

\bibitem[JVC18]{gazelle}
Chiraag Juvekar, Vinod Vaikuntanathan, and Anantha~P. Chandrakasan.
\newblock {GAZELLE:} {A} low latency framework for secure neural network
  inference.
\newblock In William Enck and Adrienne~Porter Felt, editors, {\em 27th {USENIX}
  Security Symposium, {USENIX} Security 2018, Baltimore, MD, USA, August 15-17,
  2018}, pages 1651--1669. {USENIX} Association, 2018.

\bibitem[LSI{\etalchar{+}}23]{LevySIKMZ23}
Benjamin Levy, Ben Sherman, Muhammad Ishaq, Lindsey Kennard, Ana~L. Milanova,
  and Vassilis Zikas.
\newblock Compilation and backend-independent vectorization for multi-party
  computation.
\newblock {\em {IACR} Cryptol. ePrint Arch.}, page~89, 2023.

\bibitem[LWN{\etalchar{+}}15]{oblivm}
Chang Liu, Xiao~Shaun Wang, Kartik Nayak, Yan Huang, and Elaine Shi.
\newblock Oblivm: {A} programming framework for secure computation.
\newblock In {\em 25\textsuperscript{th} {ACM} Symp.~on Operating System
  Principles (SOSP)}, pages 359--376. {IEEE}, 2015.

\bibitem[MNPS04]{fairplay}
Dahlia Malkhi, Noam Nisan, Benny Pinkas, and Yaron Sella.
\newblock Fairplay - a secure two-party computation system.
\newblock In {\em 13\textsuperscript{th} Usenix Security Symposium}, pages
  287--302, August 2004.

\bibitem[MSK23]{coyote}
Raghav Malik, Kabir Sheth, and Milind Kulkarni.
\newblock Coyote: A compiler for vectorizing encrypted arithmetic circuits.
\newblock In {\em Proceedings of the 28th ACM International Conference on
  Architectural Support for Programming Languages and Operating Systems, Volume
  3}, pages 118--133, 2023.

\bibitem[OBW22]{OzdemirBW22}
Alex Ozdemir, Fraser Brown, and Riad~S. Wahby.
\newblock {CirC}: Compiler infrastructure for proof systems, software
  verification, and more.
\newblock In {\em IEEE Symp.~on Security and Privacy}, pages 2248--2266, 2022.

\bibitem[PHGR13]{pinocchio}
Bryan Parno, Jon Howell, Craig Gentry, and Mariana Raykova.
\newblock Pinocchio: Nearly practical verifiable computation.
\newblock In {\em IEEE Symp.~on Security and Privacy}, pages 238--252. IEEE,
  2013.

\bibitem[RHH14]{wysteria}
Aseem Rastogi, Matthew~A. Hammer, and Michael Hicks.
\newblock Wysteria: {A} programming language for generic, mixed-mode multiparty
  computations.
\newblock In {\em IEEE Symp.~on Security and Privacy}, pages 655--670, May
  2014.

\bibitem[VJHH22]{heco}
Alexander Viand, Patrick Jattke, Miro Haller, and Anwar Hithnawi.
\newblock {HECO:} automatic code optimizations for efficient fully homomorphic
  encryption.
\newblock {\em CoRR}, abs/2202.01649, 2022.

\bibitem[WNW{\etalchar{+}}21]{egg}
Max Willsey, Chandrakana Nandi, Yisu~Remy Wang, Oliver Flatt, Zachary Tatlock,
  and Pavel Panchekha.
\newblock Egg: Fast and extensible equality saturation.
\newblock {\em Proc. ACM Program. Lang.}, 5(POPL), jan 2021.

\bibitem[WSR{\etalchar{+}}15]{buffet}
Riad~S. Wahby, Srinath Setty, Zuocheng Ren, Andrew~J. Blumberg, and Michael
  Walfish.
\newblock Efficient {RAM} and control flow in verifiable outsourced
  computation.
\newblock In {\em Network and Distributed System Security Symp.} The Internet
  Society, 2015.

\bibitem[Yao82]{yao82}
Andrew~C. Yao.
\newblock Protocols for secure computations.
\newblock In {\em 23\textsuperscript{rd} annual IEEE Symposium on Foundations
  of Computer Science}, pages 160--164, 1982.

\bibitem[ZCMZ03]{zcmz03}
Lantian Zheng, Stephen Chong, Andrew~C. Myers, and Steve Zdancewic.
\newblock Using replication and partitioning to build secure distributed
  systems.
\newblock In {\em IEEE Symp.~on Security and Privacy}, pages 236--250, May
  2003.

\end{thebibliography}

\end{document}